\documentclass[twocolumn,showpacs,showkeys,amsmath,amssymb,pra]{revtex4} 

\usepackage{graphicx}   
\usepackage{dcolumn}    
\usepackage{bm}         
\begin{document} 
 
\title[]{Generation of mesoscopic superpositions of a binary Bose-Einstein condensate in a slightly asymmetric double well} 
 
\author{Christoph Weiss}
\email{christoph.weiss@lkb.ens.fr}

\affiliation{Laboratoire Kastler Brossel, \'Ecole Normale Sup\'erieure, Universit\'e Pierre et Marie-Curie-Paris 6, 24 rue Lhomond, CNRS,
                F-75231 Paris Cedex 05, France
} 
\author{Niklas Teichmann}

   \affiliation{Institut Henri Poincar\'e, Centre Emile Borel, 11 rue P.\ et M.\ Curie, F-75231 Paris Cedex 05, France}

\keywords{Mesoscopic entanglement, Bose-Einstein condensation, double-well potential}
                  
\date{submitted: 27 July 2007} 
 
\begin{abstract}
A previous publication [Europhysics Letters {\bf 78}, 10009 (2007)] suggested to coherently generate mesoscopic superpositions of a two-component Bose-Einstein condensate in a double well under perfectly symmetric conditions. However, already tiny asymmetries can destroy the entanglement properties of the ground state. Nevertheless, even under more realistic conditions, the scheme is demonstrated numerically to generate mesoscopic superpositions. 
\end{abstract} 

\pacs{03.75.Gg, 03.75.Lm, 03.75.Mn}

\maketitle

\section{Introduction}

Generation of mesoscopic entanglement in a Bose-Einstein condensate (BEC) \nocite{CiracEtAl98}\nocite{SorensenEtAl01}\nocite{HolthausStenholm01}\nocite{YukalovEtAl02}\nocite{MicheliEtAl03}\nocite{MahmudEtAl03}\nocite{WeissJinasundera05}\cite{CiracEtAl98,SorensenEtAl01,HolthausStenholm01,YukalovEtAl02,MicheliEtAl03,MahmudEtAl03,WeissJinasundera05,Creffield2007} is still a challenge of fundamental research. A possible system for which entanglement could be achieved is a BEC in a double-well potential~\cite{GatiOberthaler07,MahmudEtAl03}. While attractive condensates can become unstable~\cite{DoddEtAl96}, for repulsive single species condensates the ground state is not the desired superposition~(see, e.g., Ref~\cite{HolthausStenholm01}). Nevertheless, by phase engineering fidelities (i.e.\ the probability to be in an ideal superposition) of about~$60\%$ have been obtained numerically~\cite{MahmudEtAl03}. 

In Ref.~\cite{TeichmannWeiss07} we suggested to use a binary condensate to produce Bell-like mesoscopic superpositions with fidelities above $95\%$ by using a method of tunneling control which was very recently realized experimentally in an optical lattice~\cite{LignierEtAl07}. However, to achieve the entanglement generation, the ground state of a two-component condensate of exactly the same number of particles of each kind was followed adiabatically in a perfectly symmetric double well potential~\cite{footnote07}.
Given the fact that even very small
asymmetries can destroy entanglement properties of eigen-functions~\cite{DounasfrazerCarr06}, this rises the question if the generation of mesoscopic entanglement proposed in Ref.~\cite{TeichmannWeiss07} would also be possible under more realistic conditions, i.e., slightly different particle numbers and a small tilt.

The focus of this manuscript is to investigate the scheme of entanglement generation including asymmetries in both the potential and the particle numbers. In Sec.~\ref{sec:M} the model used to describe a two-component BEC in a double well is introduced. While Sec.~\ref{sec:I} explains how the mesoscopic superpositions can be characterized, Sec.~\ref{sec:E} outlines the noticeable differences between the ideal case considered in Ref.~\cite{TeichmannWeiss07} and the more realistic (slightly asymmetric) situation considered here for which the entanglement generation no longer is entirely adiabatic.

\begin{figure}
\includegraphics[angle=0,width=\linewidth]{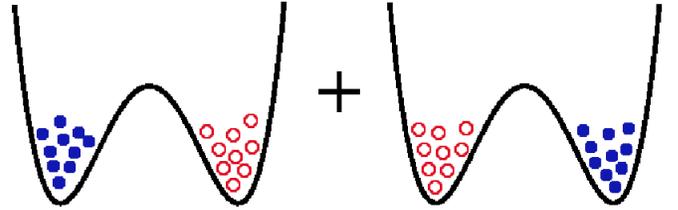}
\caption{(color online) Sketch of one of the target states for entanglement generation in a double well: a Bell-like superposition of two wave-functions for a binary Bose-Einstein condensate.}
\end{figure}

\section{\label{sec:M}Model, controlling the entanglement generation}
To describe a binary Bose-Einstein condensate in a double well,  we use the Hamiltonian in two-mode approximation (Ref.~\cite{MilburnEtAl97} cf.\ Ref.~\cite{LipkinEtAl65}):
\begin{eqnarray}
\label{eq:H}
\hat{H} &=& -\frac{\hbar\Omega}2\left(\hat{a}_1^{\phantom\dag}\hat{a}_2^{\dag}+\hat{a}_1^{\dag}\hat{a}_2^{\phantom\dag} \right) + \hbar\kappa_{\rm A}\left(\hat{a}_1^{\dag}\hat{a}_1^{\dag}\hat{a}_1^{\phantom\dag}\hat{a}_1^{\phantom\dag}+\hat{a}_2^{\dag}\hat{a}_2^{\dag}\hat{a}_2^{\phantom\dag}\hat{a}_2^{\phantom\dag}\right)\nonumber\\&& -\frac{\hbar\Omega}2\left(\hat{b}_1^{\phantom\dag}\hat{b}_2^{\dag}+\hat{b}_1^{\dag}\hat{b}_2^{\phantom\dag} \right) + \hbar\kappa_{\rm B}\left(\hat{b}_1^{\dag}\hat{b}_1^{\dag}\hat{b}_1^{\phantom\dag}\hat{b}_1^{\phantom\dag}+\hat{b}_2^{\dag}\hat{b}_2^{\dag}\hat{b}_2^{\phantom\dag}\hat{b}_2^{\phantom\dag}\right) \nonumber\\ 
&&+2\hbar\kappa_{\rm AB}\left(\hat{a}_1^{\dag}\hat{a}_1^{\phantom\dag}\hat{b}_1^{\dag}\hat{b}_1^{\phantom\dag}+\hat{a}_2^{\dag}\hat{a}_2^{\phantom\dag}\hat{b}_2^{\dag}\hat{b}_2^{\phantom\dag}\right)\nonumber\\
&&+\hbar(\varepsilon+\mu f(t))\left(\hat{a}_2^{\dag}\hat{a}_2^{\phantom\dag}-\hat{a}_1^{\dag}\hat{a}_1^{\phantom\dag}+\hat{b}_2^{\dag}\hat{b}_2^{\phantom\dag}-\hat{b}_1^{\dag}\hat{b}_1^{\phantom\dag}\right)\;,
\end{eqnarray}
where $\hat{a}^{(\dag)}_j$ ($\hat{b}^{(\dag)}_j$) creates, respectively annihilates, a boson of type A (B) in well~$j$, $\Omega$ is the tunneling frequency, $\kappa_{\rm A}$  ($\kappa_{\rm B}$) is the interaction parameter for atoms of type A (B) and  $\kappa_{\rm AB}$ models the interaction between distinguishable particles. Such Hamiltonians have also been used to describe entanglement generation beyond BEC-physics~\cite{DusuelVidal05}. The (unwanted) small tilt which might not be avoidable experimentally is represented by~$\varepsilon$.

The time-dependent potential difference $\propto f(t)$ can be used to perform tunneling control via resonant tunneling~\cite{WeissJinasundera05,Weiss06} (cf.\ Ref.~\cite{YukalovEtAl02}). References~\cite{TeichmannWeiss07,Creffield2007} generate multi-particle entanglement by 
periodic driving,
\begin{equation}
\label{eq:sin}
f(t)=\sin(\omega t)\;,
\end{equation}
to effectively renormalize the tunneling frequency~$\Omega$ by the $J_0$-Bessel function in the high frequency limit to
\begin{equation}
\label{eq:omegaeff}
\Omega_{\rm eff} = \Omega J_0(2\mu/\omega)\;.
\end{equation}
This mechanism for tunneling control (cf.~\cite{HarocheEtAl70}), which on the single particle level has been investigated theoretically already for some 16 years~\cite{GrossmannEtAl91,Holthaus92,GrifoniHanggi98,Weiss06b}, has very recently been observed experimentally for a Bose-Einstein condensate in an optical lattice~\cite{LignierEtAl07}.

Another possibility to achieve the entanglement generation of Ref.~\cite{TeichmannWeiss07} is to decrease the tunneling frequency by increasing the distance~$d$ between the two wells. 
Approximating the wave-functions of particles of mass~$m$ in the two wells by the ground states of harmonic oscillators of oscillation frequency~$\widetilde{\omega}$ and oscillator length~$\ell_{\rm osc}= \sqrt{\hbar/(2m\widetilde{\omega})}$, one obtains~\cite{HolthausStenholm01,MilburnEtAl97}
\begin{equation}
 \Omega\simeq \widetilde{\omega} \frac d{\ell_{\rm osc}}\sqrt{\frac2{\pi}}\exp\left(-\frac{d^2}{\ell_{\rm osc}^2}\right)
\end{equation}
for the tunneling frequency and
\begin{equation}
 \kappa = \frac{\hbar a}{4\sqrt{\pi}m\ell_{\rm osc}^3}\;,
\end{equation}
where $a$ is the s-wave scattering length, for the interaction parameters of Eq.~(\ref{eq:H}) as a function of $\Omega$.

A possible realization of such a model could be two hyperfine-states of, e.g, ${}^{87}$Rb. Feshbach resonances for interaction between different hyper-fine states have been observed experimentally~\cite{MarteEtAl02}, thus providing a way to slightly increase the interaction between distinguishable particles as compared to identical particles.

To coherently generate entanglement, Ref.~\cite{TeichmannWeiss07} suggests to decrease the (effective) tunneling frequency for such a situation, thus adiabatically following the ground state towards highly entangled mesoscopic states.
As decoherence-times tend to rapidly decrease with increasing particle numbers, not too large particle numbers should be used to investigate entanglement generation. Experimentally, Bose-Einstein condensates of the order of 100 particles have been realized~\cite{ChuuEtAl05}; in Ref.~\cite{TeichmannWeiss07} we used a binary BEC with exactly $N_{\rm A}=N_{\rm B}=50$ particles of each kind.  Here, deviations $N_{\rm A}=50-\Delta N/2$, $N_{\rm B}=50+\Delta N/2$ up to $|\Delta N|/(N_{\rm A}+N_{\rm B})=10\%$ will be considered.

\section{\label{sec:I}Identifying mesoscopic superpositions}
\begin{figure}
\includegraphics[angle=-90,width=\linewidth]{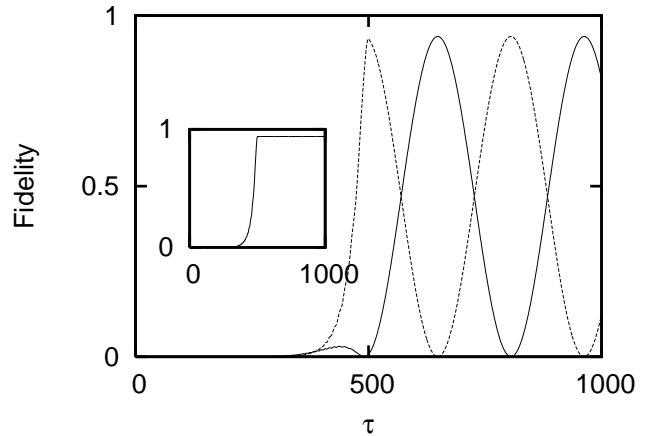}
\caption{Fidelity~(\ref{eq:proje}) as a function of dimensionless time~$\tau =\Omega t$ if the tunneling frequency is reduced linearly from $\Omega(\tau=0)=1$ to $\Omega(\tau=500)=0$ and then held at this value. Solid line: $p_+$, dashed line $p_-$.
The double well is slightly asymmetric ($\varepsilon=0.001$) with $N_{\rm A}=45$, $N_{\rm B}=55$, $\kappa_{\rm A}=\kappa_{\rm B}=0.04$; $\kappa_{\rm AB}=0.05$. \\
Inset: The same situation for an untilted double well.
Even without the tilt, the ratio $\kappa_{\rm AB}/\kappa_{\rm A }$ has to be larger than in Ref.~\cite{TeichmannWeiss07} in order to ensure that the Bell-state~$|\psi_+\rangle$ still is the ground state for a symmetric potential.
}
\label{fig:oszi}
\end{figure}

For the symmetric conditions of Ref.~\cite{TeichmannWeiss07} the fidelity obtained by projecting the wave-function~$|\psi\rangle$
on the Bell state 
\begin{equation}
\label{eq:Bell}
|\psi_{\pm}\rangle = \frac1{\sqrt{2}}\big(|0,N_{\rm A}\rangle_{\rm A}|N_{\rm B},0\rangle_{\rm B}
 \pm  |N_{\rm A},0\rangle_{\rm A}|0,N_{\rm B}\rangle_{\rm B}\big)
\end{equation}
provides an excellent measure of entanglement generation,
\begin{equation}
\label{eq:proje}
p_{\pm} = |\langle \Psi|\psi_{\pm}\rangle|^2\;,
\end{equation} 
if $p_{\pm}>0.5$. In Fig.~\ref{fig:oszi} the tunneling frequency in an asymmetric double well containing a binary BEC with $N_{\rm A}\neq N_{\rm B}$ is reduced linearly during half the time and during the other half $\Omega$ held at zero:
\begin{equation}
\label{eq:omega}
\Omega = \left\{
\begin{array}{lcr}
1-\frac{\tau}{\tau_{\rm ramp}} & : & 0\le\tau\le\tau_{\rm ramp}\\
0 & : & \tau>\tau_{\rm ramp}
\end{array}\right.\;,\quad \tau=t\Omega\;.
\end{equation}
For a slightly tilted double well, the ground state cannot be one of the two superpositions of Eq.~(\ref{eq:Bell}):
In the limit $\Omega\to 0$ the two wave-functions used in the superpositions~(\ref{eq:Bell}) do have different energies. Thus, even if a perfect superposition was reached, the phase would oscillate similar to Fig.~\ref{fig:oszi}. The fact that the fidelities stay below~$1$ indicates that more Fock states then just those included in Eq.~(\ref{eq:Bell}). Increasing the tilt by, say, a factor of ten would further increase the oscillation frequency as the energy difference between both wells increases while increasing it by a much larger factor would destroy the generation of mesoscopic superpositions. Periodic driving introduces an additional frequency, thus leading to even larger oscillations frequencies in the fidelities.

For the purpose of identifying mesoscopic superpositions independent of these oscillations, we define the modified fidelity:
\begin{equation}
\label{eq:fid}
p_{\rm fid} = \frac 12\left(\sqrt{p_1}+\sqrt{p_2}
 \right)^2\;,
\end{equation}
where
\begin{equation}
p_1=\left(\langle\Psi|0,N_{\rm A}\rangle_{\rm A}|N_{\rm B},0\rangle_{\rm B}\right)^2
\end{equation}
and
\begin{equation}
p_2=\left(\langle\Psi|N_{\rm A},0\rangle_{\rm A}|0,N_{\rm B}\rangle_{\rm B}\right)^2\;.
\end{equation}
As for the usual fidelity, this measure is~$1$ for a perfect superposition and already $p_{\rm fid}\approx 0.5$ might signify that there is no superposition at all.
The modified fidelity~(\ref{eq:fid}) can be derived by projecting the wavefunction on the superposition
\begin{equation}
\label{eq:phi}
|\psi_{\varphi}\rangle = \frac1{\sqrt{2}}\big(|0,N_{\rm A}\rangle|N_{\rm B},0\rangle
 + e^{i\varphi} |N_{\rm A},0\rangle|0,N_{\rm B}\rangle\big)
\end{equation}
and maximizing the resulting projection
\begin{equation}
\label{eq:projephi}
p_{\varphi} = \left|\langle \Psi|\psi_{\varphi}\rangle\right|^2\;
\end{equation}
as a function of $\varphi$.

\begin{figure}[t]
\includegraphics[angle=-90,width=\linewidth]{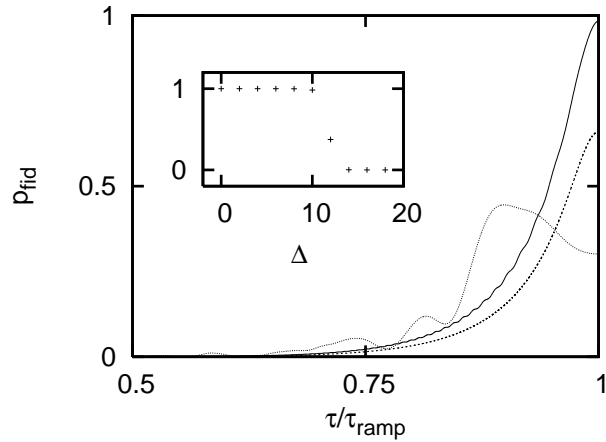}
\caption{Modified fidelity~(\ref{eq:fid}) as a function of time $t$. For ramping time $\tau_{\rm ramp}=1000$ (solid line), $p_{\rm fid}$ reaches $95.61\%$ by first adiabatically following the ground state and then non-adiabatically conserving the symmetry of the wave-function. The other parameters are chosen as in Fig~\ref{fig:oszi}.
 A ramping time of $\tau_{\rm ramp}=50$ (dotted line) is much too short as can be seen in the oscillations indicating non-adiabaticity. On the other hand $\tau_{\rm ramp}=50000$ (dashed line) is too slow; it adiabatically follows the ground state up to values where it becomes rather asymmetric. This leads to a final state with a modified fidelity well below $90\%$.\\
Inset: For a ramping time of~$\tau_{\rm ramp}=1000$, the final modified fidelity is shown as a function of the particle number difference~$\Delta$ with $N_{\rm A}=50-\Delta/2$ and $N_{\rm B}=50+\Delta/2$. Up to $\Delta /50=20\%$, the value used for most of the figures, a transfer to highly entangled states is achieved. For higher $\Delta$, the ramping-time and/or the ratio of the interaction parameters would have to be adjusted.}
\label{fig:fid}
\end{figure}

\section{\label{sec:E}Entanglement generation only partially adiabatic}
In order to understand why generation of mesoscopic superpositions is still possible, Fig.~\ref{fig:fid} shows the modified fidelity~(\ref{eq:fid}) for different ramping times~$\tau_{\rm ramp}$ (see Eq.~(\ref{eq:omega})). If the ramping is done on too short time-scales, the oscillations characteristic for too fast, i.e.\ non-adiabatic, ramping occur and high fidelities cannot be reached. If, on the other hand, the ramping is done too slowly, the system does not reach a mesoscopic superpositions either as it adiabatically follows the ground state for too long and thus becomes rather asymmetric. For moderate ramping, the system first adiabatically follows the ground state thus leading to a wave-function with a bimodal probability distribution. Then, the symmetry of the wave-function is conserved (instead of tunneling through energetically higher states with half the particles of at least one kind in both of the wells)
to finally reach a mesoscopic superposition.

\begin{figure}
\includegraphics[angle=0,width=\linewidth]{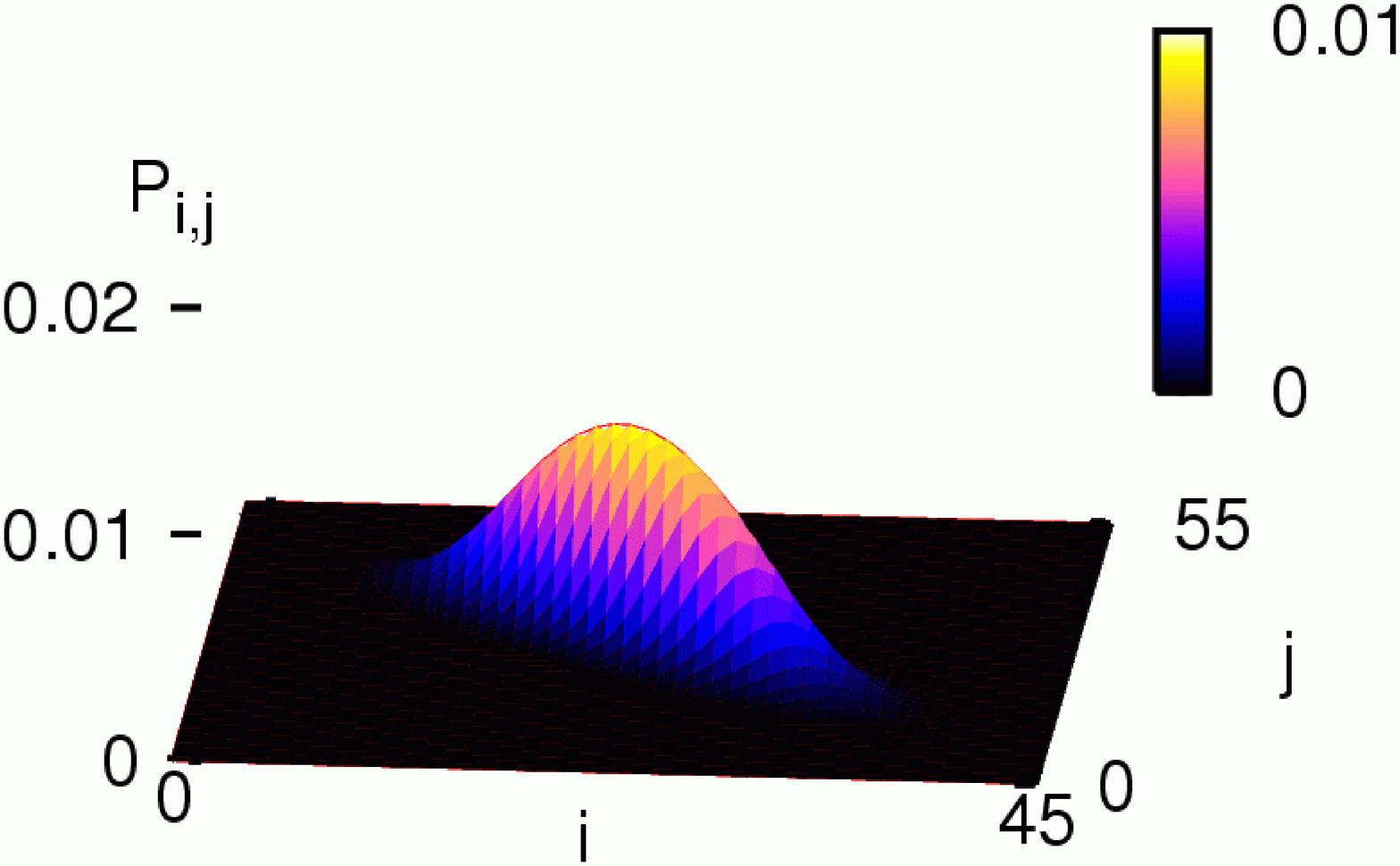}
\includegraphics[angle=0,width=\linewidth]{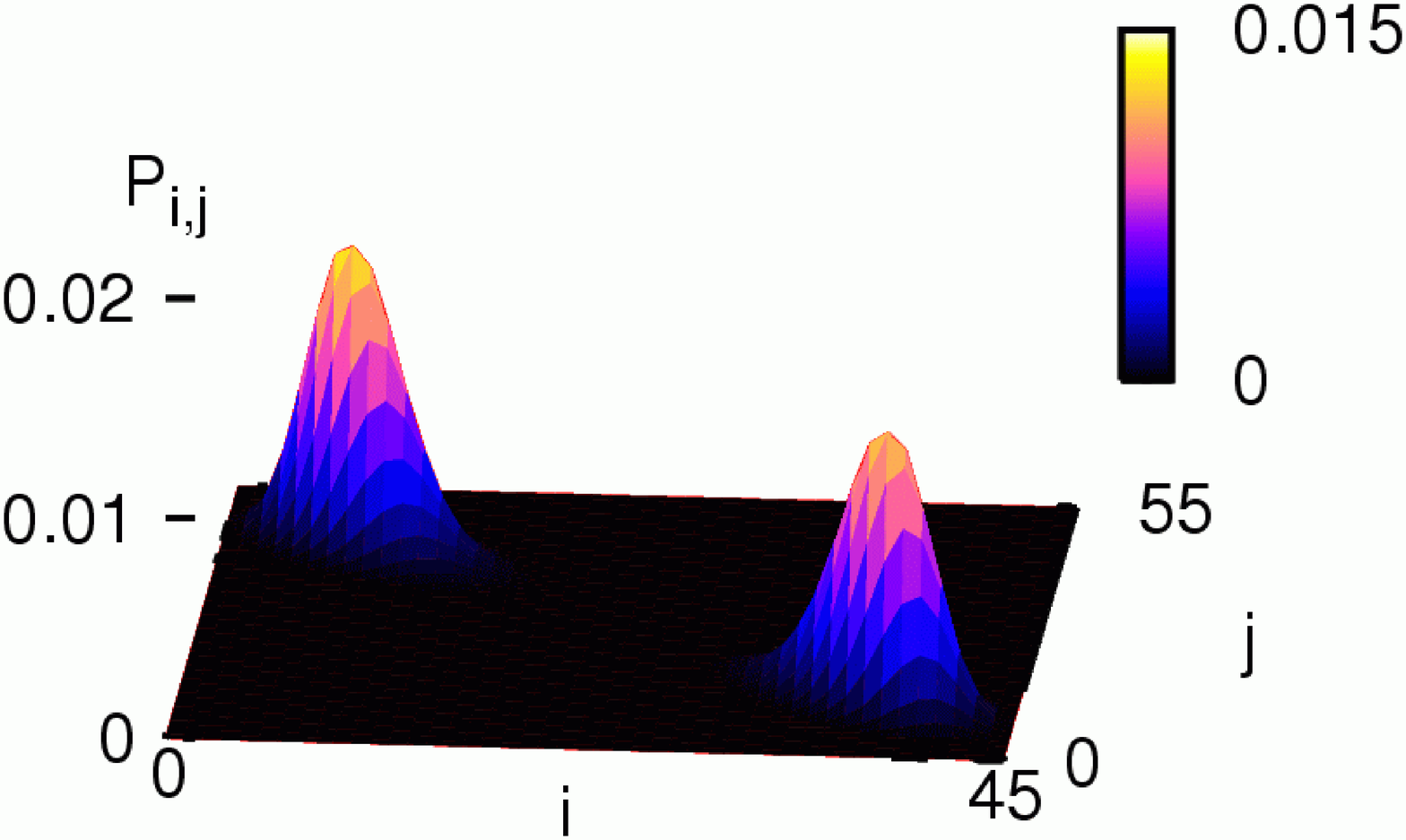}
\caption{(color online) The projection of the wave-function on Fock-states $|i,N_{\rm A}-i\rangle|j,N_{\rm B}-j\rangle$ for a binary condensate for which the tunneling control is performed via periodic driving as suggested in Ref.~\cite{TeichmannWeiss07} (see Eqs.~(\ref{eq:coh}) and (\ref{eq:omegaeff})). While the initial state at $\tau=0$ (top) is no bimodal distribution at all, a mesoscopic superposition is generated already at $\tau =400$ (bottom). The parameters where chosen as in Fig~\ref{fig:fid} with a ramping time of~$\tau_{\rm ramp}=1000$. The driving frequency $\omega=40\Omega$; at $\tau=\tau_{\rm ramp}$ the final fidelity  is again well above $90\%$.}
\label{fig:3d}
\end{figure}

If the tunneling control is done via slowly variing the amplitude~$\mu$ of periodic driving as in Ref.~\cite{TeichmannWeiss07} (cf.\ Eqs.~(\ref{eq:sin}) and (\ref{eq:omegaeff})), i.e., $\Omega=\rm const.$, $\omega\gg\Omega$ and
\begin{equation}
\label{eq:coh}
\mu\simeq \frac{2.4048\omega}2\frac{\tau}{\tau_{\rm ramp}}\;,\quad 0\le\tau\le\tau_{\rm ramp}\;,
\end{equation}
where $2.4048\ldots$ is the first zero of the $J_0$-Bessel function (see Eq.~(\ref{eq:omegaeff})),
the generation of mesoscopic entanglement also occurs. 
For the parameters in Fig.~\ref{fig:oszi} with a ramping time of $\tau_{\rm ramp}=1000$, Fig.~\ref{fig:3d} shows the emergence of an entangled wave-function from an initial state for which the probability distribution is not bimodal. 

\section{Conclusions}

While already two independent Bose-Einstein condensates provide interesting effects (see, e.g. Ref.~\cite{CastinDalibard97}), two-component BECs provide an even wider variety of effects~\cite{MyattEtAl97,BallaghEtAl97}. Here it has been shown that for a binary condensate in a double-well potential, the generation of mesoscopic entanglement suggested in Ref.~\cite{TeichmannWeiss07} works also under less ideal conditions, i.e., if the particle numbers are different and if the double well is not perfectly symmetric.

For a given experimental situation, the linear ramping used here in the spirit of coherent control theory can be optimized both toward short ramping times and large fidelities, e.g., by methods of optimal control theory~\cite{AssionEtAl98}.

\acknowledgments
We would like to thank Y.~Castin for his support. N.~T.\ thanks the Institut Henri Poincare-Centre Emile Borel
for hospitality and support.
Funding by the European Union (C.~W.\ through contract MEIF-CT-2006-038407
and N.~T.\ through contract MEST-CT-2005-019755) is gratefully acknowledged.



\begin{thebibliography}{30}
\bibitem[{foo()}]{footnote07}
\bibinfo{note}{As the coherent control was performed via time-periodic
  potential differences, strictly speaking the corresponding Floquet-state was
  followed adiabatically, see also~\cite{BreuerHolthaus89}}.
\end{thebibliography}
\end{document}